\documentstyle[11pt]{article}

\newcommand{\be}{\begin{equation}}
\newcommand{\ee}{\end{equation}}
\newcommand{\ba}{\begin{array}}
\newcommand{\ea}{\end{array}}
\newcommand{\bc}{\begin{center}}
\newcommand{\ec}{\end{center}}
\newcommand{\bi}{\begin{itemize}}
\newcommand{\ei}{\end{itemize}}

\newcommand{\disregard}[1]{{}}

\def\bild#1\over#2{\mathrel{\mathop{\kern0pt #1}\limits_{#2}}}
\begin{document}

\centerline {{\bf RANDOM MAGNETIC IMPURITIES AND THE LANDAU PROBLEM  \rm}}
\vskip 2cm
\centerline{{\bf Jean DESBOIS, Cyril FURTLEHNER\rm } and
{\bf St\a'ephane OUVRY \rm }\footnote{\it  and
LPTPE, Tour 12, Universit\'e Paris  6 / electronic e-mail:
OUVRY@FRCPN11.IN2P3.FR}}
\vskip 1cm
\centerline{{Division de Physique Th\'eorique \footnote{\it Unit\a'e de
Recherche  des
Universit\a'es Paris 11 et Paris 6 associ\a'ee au CNRS},  IPN,
  Orsay Fr-91406}}
\vskip 3cm
{\bf Abstract : The 2-dimensional density of states of
an electron is studied for a Poissonian random distribution of point
vortices carrying $\alpha$ flux in unit of the quantum of flux.
It is shown that, for any given density of impurities, there is a
transition, when $\alpha\simeq 0.3-0.4$, from an "almost  free" density of
state -with only a
depletion of states at the bottom of the spectrum characterized by a
Lifschitz tail- to a Landau density of state with sharp Landau level
oscillations. Several evidences and arguments for this
transition -numerical and analytical-
are presented.}

\vskip 0.5cm


IPNO/TH 95-46

June 1995

\vfill\eject

{\bf I. Introduction :}

One considers a 2-dimensional model for an electron of electric
charge $e$ and of mass $m$
subject to  a random magnetic field.
Here, random magnetic field means a Poissonian\footnote{ One could
consider as well other types of probability distribution - Gaussian for
example. For the sake of simplicity, we concentrate only on the Poisson
case.} random distribution $\{\vec r_i, i=1,2,....,N\}$
of fixed infinitely thin vortices carrying a flux
$\phi$, modeling magnetic impurities, and characterized  by
the dimensionless Aharonov-Bohm ($A-B$) coupling
$\alpha=e\phi/2\pi=\phi/\phi_o$.

The question we ask is about the effect of disorder on the energy level density
of
an electron -test particle-
averaged over the random position of the vortices \cite{0}.
In the thermodynamic limit  $N\to
\infty, V\to\infty$ for a distribution of vortices of
density $\rho=N/V$, one might naively argue that the average magnetic field,
$<B>=\alpha \rho\phi_o$, is expected to become meaningful
in the limit $\rho\to \infty,\alpha\to 0$, with
$\rho\alpha$ kept finite. On the other hand, if $\rho$ is
finite, and $\alpha$ non vanishing,
corrections due to
disorder should  exhibit non trivial magnetic
impurity signatures, like broadening of Landau levels and localization.

In a first paper \cite{DFO} on this problem, two approaches were used:

- A path integral Brownian motion analysis
where the problem at hand is mapped, after averaging on disorder, to a
study of winding properties of Brownian curves.

- A quantum mechanical formulation, where the contact hard-core
boundary conditions at the location of the impurities where properly taken into
account by an
appropriate
wave function redefinition, allowing for an analytical averaging on
the random singular $A-B$ interactions.

The main results were :

- in the presence of a constant external magnetic field,
if  the total (external + average)
magnetic field  is strong enough so that one can neglect
the coupling between
the lowest Landau level and the excited Landau levels by
the random component of the vortex distribution,  the
system was best described when projected in the LLL.
Since one  has in view a sufficiently
dilute gas of electrons compared to the  available
quantum states in the LLL -the fractional Quantum Hall
regime-, such a restriction is licit. In this situation,
the quantum problem was explicitly
mapped on a problem of random $\delta$ impurities, where the average
density of states happened to be known \cite{bre}.

- in the average magnetic field limit $\alpha\to0,
\rho\to\infty$,
a global zero point energy shift ${|e<B>|\over 2m} = <\omega_c>$
materialized
in the Landau spectrum of the
average magnetic field $<B>$. The origin of this shift was traced back to the
hard-core
boundary
conditions at the location of the impurities \cite{hard}.

- in the case $\alpha= 1/2$ and $\rho$ finite,
on the other hand, Brownian
numerical
simulations showed that
the sole effect of the impurities was
a depletion of states at the bottom of the spectrum,
 characterized by a Lifschitz tail in the average density of states, caused by
the
isolated  impurities.

An additional interesting result was that the average
density of states
$<\rho(E,\alpha,\rho)>$ happened to be a function of $E/\rho$ and
$\alpha$ only (this is also true in presence of an external magnetic
field). The $<\rho(E/\rho,\alpha)>$ scaling implies that  the
impurity density $\rho$ is not a relevant
parameter, and can be arbitrarily set to a given
value, since changing its value amounts simply to a rescaling of the energy
unit. A consequence is a more precise definition of the limit
where the average magnetic field becomes meaningful. {\bf It should
happen,
for any given finite $\rho$, when $\alpha$ becomes sufficiently small,
independently of the value of $\rho$. Thus  one expects a critical value
$\alpha_c$ where a transition occurs from an almost free density of
states with a
Lifschitz tail at the bottom of the spectrum ($\alpha_c<\alpha<1/2$),
to a Landau like
density of states, with Landau oscillations, i.e. Landau
 levels separated by a Landau gap
($0<\alpha<\alpha_c$).}

A
semiclassical understanding of this transition consists in taking an
electron with a Fermi velocity $v_F$, subject to the average magnetic field
$<B>=\rho\alpha\phi_o$. A
typical cyclotron orbit radius
is $R\simeq v_F/(\rho\alpha)$. For a given
$\rho$, one has $R\simeq 1/\alpha$, thus the smaller is $\alpha$, the larger is
the
number of magnetic impurities enclosed by the cyclotron orbit, and
therefore the
more accurate is description in terms of the average magnetic field. To
summarize, for a
given finite $\rho$, the smaller is  $\alpha$, the better
is $<B>$, eventhough it is smaller and smaller.
However,  there is no other magnetic scale
in the problem to compare it with.

An important consequence of the Lifschitz tail - Landau levels transition
should be
a non conducting-conducting transition, since one expects localization due
to disorder when $\alpha>\alpha_c$, and extended states in the
Landau regime when $\alpha<\alpha_c$.

The aim of the present paper is to show that the critical value at which the
transition occurs
is $\alpha_c\simeq 0.3-0.4$.

We will present two types of evidence for this result :

- Numerical evidences

i) for the specific heat, where a Brownian motion numerical study indicates a
transition from an almost  free type to a Landau type specific heat when
$\alpha_o^{num}\simeq 0.28$.

ii) for the density of state, where, under some reasonnable
simplifying assumptions, the
transition is shown to occur at $\alpha_c\simeq 0.35$.

- Analytical evidences :

i) one can explicitly show that
 the specific heat transition is possible only when Landau
oscillations have already begun, implying that
 $\alpha_c> \alpha_o$, but close to $\alpha_o$.

ii) quantum mechanical evidence using the impurity cluster expansion of the
average partition function for an arbitrary large number of impurities.
At a given order $\rho^N$,
for a test particle subject to a given number $N$ of impurities described by
the
Hamiltonian $H_N$, the average partition function $<Tr \exp(-\beta H_N)-Tr
\exp(-\beta H_0)>$ has been computed as a power series in $\alpha$.
One recovers, at order $\rho^N\alpha^N$, the leading order in
$\alpha$, the partition function of the average
magnetic field.
Higher order corrections can be computed : for
$N=2$, the diagrammatic expansion has been done up to order $\alpha^4$.
We will show that this is all what is
needed to obtain
$\alpha_o\simeq 0.29$, which is indeed very close to the numerical
result $\alpha_o^{num}=0.28$.

{\bf II. Definition of the model :}

The system is peridodic in
$\alpha$ with period 1, so one can always take $\alpha\in[0,1]$.
In
the absence of an external magnetic field,
there is no privileged orientation to the plane, therefore the system is
symetric
wrt $\alpha=1/2$, and one can restrict
$\alpha\in [0,1/2]$. It follows that any physical quantity of interest
should depend
on $\alpha(1-\alpha)$ only.

{\bf II.a  Path integral Brownian approach :}

One starts \cite{DFO} from a square lattice of $\cal{N}$ squares of size
$a^2$, in which point magnetic impurities  are randomly dropped. Let $N_i$ be
the number
of vortices dropped on square $i$. A random configuration $\{N_i\}$ will be
realized with the probability
\be\label{prob} P(\{N_i\})={{N}!\over
{\cal{N}}^{{N}}\prod_{i=1}^{{\cal{N}}} N_i!}\to_{{\cal{N}}\to\infty}
\prod_i^{{\cal{N}}}
{(\rho a^2)^{N_i}e^{-\rho a^2}\over N_i !}\ee
with ${N}/{\cal{N}}=\sum_{i=1}^{{\cal{N}}} N_i/{\cal{N}}=\rho a^2$. In
the thermodynamic limit, this is nothing but a Poissonian distribution.
In order to compute the average level density $<\rho(E)>$, one focuses,
in the thermodynamic limit $\cal{N}\to \infty$, on the one-electron
average partition function per unit volume (for an electron of unit mass
and charge)
\be\label{PART} Z=Z_o <e^{i\sum_{i=1}^{\cal{N}}  2\pi n_i
N_i\alpha}>_{\{C,N_i\}}\ee
where $\{C\}$ is the set of $L$ steps closed random
walks, and
 $n_i$ is the number of times the square $i$ has been wound around by a
given
random walk in $\{C\}$, i.e. its winding number. $Z_o={1\over 2\pi
t}$ is the free partition
function per unit volume, with $t$ (the inverse temperature $\beta$) the length
of the curve
($2t=L a^2, e=m=1$).  (\ref{PART}) is obviously
invariant when $\alpha$ is shifted by an
integer, and when $\alpha\to-\alpha$ (because $n_i$ comes always
with $-n_i$),
so one can always restrict to $0<\alpha<1/2$.
Averaging $Z$ with (\ref{prob}) one gets
\be\label{PART'} Z=Z_o<e^{\rho\sum_nS_n(e^{i 2\pi \alpha n}-1)}>_{\{C\}}\ee
where $S_n$ stands for the arithmetic area of the $n$-winding sector
\cite{cdo} of
a given random walk  in $\{C\}$. Eq. (\ref{PART'}) is still true in the limit
$a\to 0, L\to \infty$, with
$t$ fixed, i.e. for Brownian curves in the plane, yielding a path integral
formulation for the problem at hand.
Extracting the variable $t$, it rewrites as
\be \label{SA}
Z=Z_o\int e^{-\rho t(S+iA)}P(S,A)dSdA\equiv Z_o<e^{-\rho t(S+iA)}>_{\{C\}}
\ee
where ${S}$ and ${A}$ are defined as
\be \label{sum} {{S}}={2\over t}\sum_nS_n\sin^2(\pi\alpha
n); \quad <{{S}}>={\pi\alpha}(1-\alpha) \ee
\be {A}={1\over t}\sum_nS_n \sin(2\pi\alpha n); \quad  <A>=0 \ee
$P(S,A)$ is the joint probability distribution of the random
variables $A$ and $S$.
One has used the general property of Brownian curves  that $S_n$ scales like
$t$ (remind that $<S_n>=t/(2\pi n^2)$ \cite{cdo}), so that the
variables $S$ and $A$ are actually $t$ independent.
This implies that the average partition function $Z$ in (\ref{SA})
 has the form ${F(\rho t)/t} $, and thus
its inverse Laplace
transform, the average density of states,
is necessarily a function of $E/\rho$
and $\alpha$.

{\bf II.b Microscopic Quantum Hamiltonian :}

The Hamiltonian  of an electron of mass $m$ and charge $e$
subject to the potential vector of $N$ vortices at position $\vec r_i$
($\vec k$ is the unit vector perpendicular to the plane)
reads \cite{DFO}
\be \label{H} H_N={1\over 2m}
\left(\vec p - \sum_{i=1}^N\alpha{\vec k\times(\vec r-\vec r_i)
\over (\vec r -\vec r_i)^2}
\right)^2
\ee
Because of
periodicity in $\alpha$,  one can always restrict $\alpha$ to
$\alpha\in [-1/2,+1/2]$.
We consider  in the thermodynamic limit the Poisson probability distribution
$dP(\vec r_i)={d\vec r_i/V}$.

The system described by the
Hamiltonian (\ref{H}) is not yet entirely defined. Boundary conditions
on the wave functions
have to be specified when the electron comes close to an impurity. This
can be achieved in a non ambiguous way by  adding to the pure Aharonov-Bohm
Hamiltonian
the contact terms
\be H_N^{\pm}={1\over 2m}
\left(\vec p - \sum_{i=1}^N\alpha{\vec k\times(\vec r-\vec r_i)
\over (\vec r -\vec r_i)^2}
\right)^2
\pm \sum_i {\pi|\alpha|\over m}\delta^2(\vec r-\vec r_i)
\ee
These contact terms
amount to couple the
spin-up ($+$) or spin-down ($-$) degree of freedom \cite{stefan} of the
electron
endowed with a magnetic moment
$\mu=-{e\over 2m}\alpha/|\alpha|$ (thus an electron with gyromagnetic
factor $g=2$),  to the infinite magnetic field
inside the flux-tubes.
Their origin can also be understood in a perturbative framework : the
original Aharonov-Bohm spectrum with vanishing wavefunction  at the location of
the magnetic impurities (a particular self adjoint extension describing
impenetrable vortices : hard-core)
can be perturbatively obtained if and only if
the contact terms with the ($+$) sign are taken into account, whereas
the contact terms with the opposite ($-$) sign
correspond to a different self-adjoint extension, where the
the wavefunction is
singular at the location of the vortices (the
particle is attracted inside the vortices : attractive-core).
 Note that in the Brownian path integral formulation, only the
hard-core case can be described, due to the fact that, by definition,
a given Brownian
path has no chance to pass through a given impurity location.

The contact terms happen to be crucial for the averaging on the
disorder : consider the nonunitary wavefunction redefinition \cite{acad}
\be \label{new}\psi_N^{\pm}(\vec r)
=
\prod_{i=1}^N\vert\vec r-\vec r_i\vert^{\pm |\alpha}|
\tilde{\psi}_N^{\pm}(\vec r)\ee
to obtain the Hamiltonian $\tilde{H}_N^{\pm}$ acting on
$\tilde{\psi}_N^{\pm}(\vec r)$
where the impurity potential now reads
\be\label{100} \tilde{H}_N^{\pm}=-{2\over m}\partial_{\bar z}\partial_z
-{\alpha \pm |\alpha|\over m}\sum_{i=1}^N{\partial_z
\over\bar z-\bar z_i}
+{\alpha \mp |\alpha|\over m}\sum_{i=1}^N{\partial_{\bar z}
\over z- z_i}\ee
which rewrites if $\alpha>0$ as
\be \label{101} \tilde{H}_N^{+}=-{2\over m}\partial_{\bar z}\partial_z
-{2\alpha \over m}\sum_{i=1}^N{\partial_z
\over\bar z-\bar z_i}
\ee
\be \label{1010} \tilde{H}_N^{-}=-{2\over m}\partial_{\bar z}\partial_z
+{2\alpha \over m}\sum_{i=1}^N{\partial_{\bar z}
\over z-z_i}
\ee
and if $\alpha<0$
\be \label{102}
\tilde{H}_N^{+}=-{2\over m}\partial_{\bar z}\partial_z
+{2\alpha \over m}\sum_{i=1}^N{\partial_{\bar z}
\over z- z_i}
\ee
\be \label{1020} \tilde{H}_N^{-}=-{2\over m}\partial_{\bar z}\partial_z
-{2\alpha \over m}\sum_{i=1}^N{\partial_z
\over\bar z-\bar z_i}
\ee
One sees in both $\tilde H_{N}^{\pm}$ cases that there
is again no distinction to
be made between $\alpha\in [0,1/2]$ and $\alpha\in[-1/2,0]$.
Indeed, the impurity potential are hermitian conjugate the one from the
other, but perturbative computations in $\alpha$ are always such that real
results are obtained. So one can always restrict to $\alpha\in [0,1/2]$.
Both $\tilde H_{N}^{\pm}$
cases should be considered in principle. However, it is easy to see that
they are equivalent, except for a
global shift $+<\omega_c>$ in the hard-core
$\tilde H_{N}^{+}$ case, and
$-<\omega_c>$ in the   attractive-core $\tilde H_{N}^{-}$ case. Indeed,
consider, instead of (\ref{new}), the wavefunction redefinition where the
average magnetic field Landau pre-exponential factor has been extracted
\cite{DFO}
\be \label{neww}\psi_N^{\pm}(\vec r)
=e^{\mp{1\over2}m<\omega_c>r^2}
\prod_{i=1}^N\vert\vec r-\vec r_i\vert^{\pm \alpha}
\tilde{\psi}_N^{\pm}(\vec r)\ee
to get (in the thermodynamic limit $N\to\infty$)
\be\label{extract} \tilde{H}_N^{\pm}=\pm<\omega_c> +H_{\pm
<B>}+V^{\pm}(\alpha)-<V^{\pm}(\alpha)>\ee
where $H_{<B>}$ is the Landau Hamiltonian for the average $<B>$ field
and the impurity potential reads
\be V^{+}(\alpha)-<V^{+}(\alpha)>=    -{2\alpha\over
m}\sum_{i=1}^N{\partial_z\over \bar z-\bar z_i}+2<\omega_c>z\partial_z+
\sum_{i=1}^N<\omega_c>\alpha{{\bar z}\over {\bar
z}-{\bar z_i}}-m<\omega_c>^2{\bar z} z\ee
($V^{-}(\alpha)-<V^{-}(\alpha)>$ is obtained by taking the hermitian conjugate
of
$V^{+}(\alpha)-<V^{+}(\alpha)>$ and $\alpha\to-\alpha$).
Clearly, in the limit with no disorder $\alpha\to 0$, one obtains
the Landau Hamiltonian for the average $<B>$ field
with a $\pm<\omega_c>$ shift.

The Hamiltonians $H_{N}^{\pm}$ and $\tilde H_{N}^{\pm}$ are equivalent, and can
be
indifferently used for computing the partition
function or the density of states. However,
interactions with
two magnetic
impurities have disappeared from $\tilde{H}_N^{\pm}$,
greatly simplifying the average on the disorder, which
can be easily done
using the identities
\begin{eqnarray} \int d\bar z_idz_i {1\over \bar z-\bar
z_i}\partial_{z}&=&\pi
z\partial_z\\
\int d\bar z_idz_i{1\over \bar z-\bar z_i}\partial_{z}
{1\over \bar z'-\bar z_i}\partial_{z'}&=&\pi ({z\over \bar z'-\bar z}
+{z'\over \bar z-\bar z'})
\partial_{z}\partial_{z'}\\
\int d\bar z_idz_i{1\over \bar z-\bar z_i}\partial_z
{1\over \bar z'-\bar z_i}\partial_{z'}
{1\over \bar z"-\bar z_i}\partial_{z"} &=& \nonumber \\
 \pi ({ z\over (\bar z'-\bar z)(\bar z"-\bar z)}+
{ z'\over (\bar z"-\bar z')(\bar z-\bar z')}&+&
{ z"\over (\bar z'-\bar z")(\bar z-\bar z")})\partial_{z}\partial_{z'}
\partial_{z"}\end{eqnarray}
etc....

Simple dimensional arguments at the level of the Hamiltonian are
sufficient to understand the scaling
property of the average density of states.
Rescaling the length unit by $\lambda$ amounts to rescale the
Hamiltonian, thus the energy, by $1/\lambda^2$. On the other hand, in
$d=2$, the same length unit rescaling implies for the density $\rho$ the same
$1/\lambda^2$ rescaling. It is not surprising to find, after averaging, the
$E/\rho$ scaling of the density of states.

In the sequel one will concentrate only on the hard-core case, bearing
in mind that the attractive-core case can be straightforwardly deduced
from the hard-core case.

{\bf II.c Known results: the cases $\alpha\to 0$ and $\alpha=1/2$}

To simplify notations, we will use from now on the Brownian notations
$e=m=1$, and
$\beta=t$.

{}From \cite{DFO}   we know that :

i) In the limit $\alpha\to 0$, one expects from (\ref{SA}) that
$Z\to Z_{<B>}=Z_o<e^{i<B>\sum_n nS_n}>_{\{C\}}$, the partition function of one
electron in an uniform magnetic field
$<B>=\alpha\rho\phi_o$.
However, possible corrections coming from the exponent $\exp(i2\pi n\alpha)-1$
might alter this result. Due to the non-differentiability of
Brownian paths,
$\sum_n
n^2S_n$ is not defined for a typical Brownian curve where
$<S_n>={t/(2\pi n^2)}$ \cite{cdo}. Certain
recent results in the mathematical litterature \cite{werner} show that, for
$n$ sufficiently large, $n^2S_n\to <n^2S_n>=t/(2\pi)$. It follows that,
when $\alpha\to 0$, $S\simeq <S>\simeq \pi\alpha$, and $A\simeq
2\pi\alpha{\cal A}$, where ${\cal A}$ is the algebraic area enclosed by
the Brownian curve (${\cal A}=\sum nS_n/t$). One deduces that
$Z\to_{\alpha\to 0} Z_{<B>}e^{-t<B>/2}$,
implying that the system of random vortices is equivalent to an uniform
magnetic field $<B>$, but with an additional positive shift
in the Landau
spectrum
$<B>/2=<\omega_c>$. Note that $Z_{<B>}$ is built by the
random variable ${ A}$, and the shiff by the random variable $S$.

ii) When $\alpha= 1/2$ on the other hand, one can explicitly test the effect of
the random distribution of vortices.
(\ref{PART'}) now reads
$ Z=Z_o<e^{-\rho t S}>_{\{C\}}$
and the average density of states, obtained by inverse
Laplace transform of $Z$, is
\be \label{rho}<\rho(E)>=\rho_o(E)\int_0^{{E\over \rho}}P(S)dS\ee
$P(S)$, the probability distribution for the random variable $S$, was
estimated numerically
by simulations on a lattice, where a number of steps ranging from 2000 to
32000 was used.
In Figure 1, $<\rho(E)>$ displays
 a Lifschitz tail at the bottom of the spectrum, around
$E\simeq \rho<S>=\pi\rho/4$, where a behavior
$<\rho(E)>\simeq \exp(-\rho/E)$ is expected.

In both these extreme cases $\alpha=0,\alpha=1/2$, an energy level depletion
at the bottom of the spectrum is  observed (global shift in the Landau
spectrum in the former case, Lifschitz tail in the latter).
This pattern is quite reminiscent of one impurity $A-B$ density of states
depletion $\rho_1(E)-\rho_o(E)={
\alpha(\alpha-1)\over 2}\delta(E)$ \cite{deplet}, which occurs precisely
at $E=0$.

The question we now ask is what happens when $\alpha$ continously
decreases from $\alpha=1/2$ to $\alpha=0$?
In particular, is it possible to understand the transition from a
Lifschitz tail pattern to a Landau pattern in terms of several
impurities average density of states?

{\bf III. Numerical evidences}

{\bf III.a Specific heat }

Interesting enough is the study of the specific heat averaged over
disorder
\be\label{heat} c=kt^2{d^2\over dt^2}<\ln Z'>_{\{N_i\}}\ee
where $k$ is the Boltzman constant and
\be Z'=Z_o <e^{i\sum 2\pi n_iN_i\alpha}>_{\{C\}}\ee
is the partition function
for an electron in a given distribution of vortices $\{N_i\}$. We first
note  that
\be <\ln Z'>_{\{N_i\}}=\ln[<Z'>_{\{N_i\}}](\equiv \ln Z)\ee
In principle this property holds only for short ranged impurity
potentials.
This is of course not the case in the present problem. However, before
averaging over disorder, the partition function involves only closed
Brownian curves, and thus is entirely determined by the impurities
distributed inside each Brownian curve (see (\ref{PART})). Using the basic
property that two Brownian curves on the plane have no chance to
intersect each other, it is easy to get
\be <Z'^2>_{\{N_i\}}=<Z'>^2_{\{N_i\}}\ee
i.e. $Z'$ is a self-averaging quantity.
A straightforward computation leads to
the $t$ (inverse temperature) expansion (note that
$P(S,A)=P(S,-A)$ )
\be c\simeq c_o+kt^2(<S^2>_{\{C\}}-
<S>^2_{\{C\}}-<A^2>_{\{C\}})+\cdots\ee

Both quantities $<S^2>_{\{C\}}-
<S>^2_{\{C\}}$ and $<A^2>_{\{C\}}$ have a natural interpretation in
terms of a Landau spectrum : $<A^2>_{\{C\}}$ determines the average magnetic
field Landau levels, and
 $<S^2>_{\{C\}}-
<S>^2_{\{C\}}$ measures the deviation from the Landau pattern due to
disorder. It is not a surprise that their difference
$(<S^2>_{\{C\}}-
<S>^2_{\{C\}})-<A^2>_{\{C\}}$
plays a
role in finding the critical point for the specific heat. They have
been studied
numerically, and are displayed in Fig. 2 for 2000 random walks of length
$L=100000$. When $\alpha$ continously decreases from $\alpha=0.5$ to
0, $c-c_o$ is first positive, then vanishes for $\alpha\simeq 0.28$, then
becomes
negative. The vanishing of the $t^2$ term in the specific heat
corresponds to a perfect gas behaviour. We will argue below, by
analytical means, that the perfect gas behaviour takes place at
$\alpha_o\simeq 0.29$.

{\bf III.b The density of states :}

In order to estimate $<\rho(E)>$ numerically, a simple heuristic ansatz
has to be made.
Correlators between the variables $S,A$ and
the algebraic area ${\cal A}\equiv \sum_nnS_n/t$, such  as
\be C_{S,A}\equiv {<SA>-<S><A>\over \sqrt{(<S^2>-<S>^2)(<A^2>-<A>^2)}}\ee
or $C_{S,{\cal A }}$ are trivially vanishing because of obvious
symmetry properties, and carry no information. It is more
appropriate to use the variables $|A|$, and $|{\cal A}|$. Numerical
simulations indicate that $C_{|A|,|{\cal A}|}\simeq 1$ for $\alpha<0.05$,
and $C_{|A|,|{\cal A}|}\simeq 0.93-0.95$ for $\alpha\simeq 0.25$.
One deduces that the variables $|A|$ and $|{\cal A}|$ are highly correlated, in
particular when $\alpha$ is very small, i.e. in the average magnetic
field limit.
We thus assume the linear relation
\be\label{assume} |A|={B\over \rho}|{\cal A}|\ee
where $B$ is an effective magnetic field determined by
$B/\rho=\sqrt{{<A^2>\over <{\cal A}^2>}}=\sqrt{12 <A^2>}$, which gives,
in the limit $\alpha\to 0$ that $B=2\pi\alpha\rho=
<B>$, as it should\footnote{To obtain $<{\cal A}^2>=1/12$, consider
simply the probability distribution for the random variable ${\cal A}$,
which is nothing but the Fourier transform of the partition function of
a charged particle in a constant magnetic field  \cite{cdo}
\be P({\cal A})= {1\over 2\pi}\int_{-\infty}^{+\infty}dBe^{iB{\cal
A}}{B \over 2\sinh B/2}\ee}. On
the other hand, in the limit $\alpha \to 1/2$, $B=0$, the magnetic
field is absent, as again we already know from  (\ref{SA}).
In conclusion, the ansatz (\ref{assume}) seems quite
reasonable. Now, introduce the positive random variable
\be
S'=S-C_{S,|{\cal A}|}\sqrt{{<S^2>-<S>^2\over<{\cal A}^2>-<|{\cal
A}|>^2}}|{\cal A}|\ee
so that $S'$ and $|{\cal A}|$ are uncorrelated. The average partition function
rewrites as
\be Z\simeq Z_o<e^{-\rho tS'}><e^{\rho t(S'-S)}\cos(tB{\cal A})>\ee
The inverse Laplace transform of $Z/<e^{-\rho tS'}>$ can be obtained
analytically
\be \rho'(E')=\rho_o(E'){1-e^{-2u}\over 1+e^{-2u}+2e^{-u}\cos \gamma}\ee
where
\be \gamma=2\pi{(E'/\rho)(B/\rho)\over\mu^2+(B/\rho)^2}\ee
and
\be u={\gamma\mu\over (B/\rho)}\ee
where $\mu=  C_{S,|{\cal A}|} \sqrt{{<S^2>-<S>^2\over<{\cal A}^2>-<|{\cal
A}|>^2}}$.
The average density of states rewrites
\be<\rho(E)>=\int_0^{E/\rho} P(S')\rho'(E-\rho S')dS'\ee
where $P(S')$ is the probability distribution for the variable $S'$.
The numerical results are displayed in Fig. 3  for different values of
$\alpha$. It is clear that the average density of states is no more
monotonic when $\alpha$ is below  the critical value $\alpha_c\simeq
0.35$. Also, when $\alpha\to 0$, well separated Landau peaks appear with
a well defined Landau gap. In this low energy region (which corresponds
to long Brownian curves), the system clearly mimicks a constant average
magnetic field.

{\bf IV. Analytical evidences}

{\bf IV.a $\alpha_c > \alpha_o$ :}

Let us show that the transition for the density of states
occurs necessarily at a critical $\alpha_c>\alpha_o$. Let us assume
that when $E\to\infty$, $<\rho(E)>\to \rho_o(E)={V\over 2\pi}$, i.e. at very
high
energy the system does not see the magnetic impurities, and when
$E\to 0$, $<\rho(E)>\to 0$, i.e. at very small energy the depletion
of states due to the hard-core impurities is effective. Both these
assumptions are quite reasonnable, and are actually verified in
all the numerical simulations and analytical studies. After integration by part
in $Z$, one
gets
\be c=c_o+{kt^2\over
2}{\int_0^{\infty}\int_0^{\infty}dEdE'e^{-t(E+E')}{d<\rho(E)/V>\over dE}
{d<\rho(E')/V>\over dE'}(E-E')^2\over
\{\int_0^{\infty}e^{-tE}{d<\rho(E)/V>\over dE}\}^2}\ee
At small $t$, this expression becomes
\be c\simeq c_o+kt^22\pi^2
\int_0^{\infty}\int_0^{\infty}dEdE'{d<\rho(E)/V>\over dE}{d<\rho(E')/V>\over
dE'}(E-E')^2\ee
which in turn implies that for $c-c_o$ to change its sign one necessarily
should already have Landau oscillations.

{\bf IV.b The concentration expansion :}

In the thermodynamic limit, the number of impurities is only fixed on
average. We are in a situation similar to the one encountered in
statistical mechanics for an undetermined number of particles, where
there exists a cluster expansion for the thermodynamical potential,
which involves at a given order of the fugacity $z^N$, the partition
functions for $N$ particles, $N-1$ particles, etc...
Here, there is one particle subject to the effect of an undetermined
number of impurities. Again, there exists an expansion for
the averaged one
particle partition function \cite{pastur}, analog to the
cluster expansion. The parameter  of the expansion is the
density (concentration) of impurities $\rho$, and at a given order $\rho^N$,
the
partition function per unit volume is expressed in terms of the
partition
functions for
$N$ impurities, $N-1$ impurities, etc... It reads
\begin{eqnarray} \label{conc} Z&=&Z_0(1+\rho V({Z_1\over Z_0}-1)+{1\over2
!}\rho^2V^2[{Z_2\over Z_0}-2{Z_1\over Z_0}+1]
+{1\over 3!}\rho^3V^3[{Z_3\over Z_0}-3{Z_2\over Z_0}+3{Z_1\over
Z_0}-1]\nonumber\\
&+&{1\over 4!}\rho^4V^4[{Z_4\over Z_o}-4{Z_3\over Z_o}+6{Z_2\over
Z_o}-4{Z_1\over Z_o}+1]+...\end{eqnarray}
where $Z_N\equiv <Tr \exp(-t H_N)>/V$.

Let us compare (\ref{conc}) with the $\rho t$ expansion of (4), namely
\be \label{t} Z=Z_o(1-\rho t<S>+{1\over 2!}\rho^2t^2<(S+iA)^2>-{1\over
3!}\rho^3t^3
<(S+iA)^3>+\cdots)\ee
So, considering $\rho$ as the expansion parameter, one sees that
in order for both expansions (\ref{conc}) and (\ref{t}) to match
one necessarily has
\be \label{relat}{Z_1\over Z_0}-1=-{t\over V}<S>,\quad {Z_2\over
Z_0}-1-2({Z_1\over
Z_0}-1)={t^2\over V^2}<(S+iA)^2>, \cdots\ee

One already knows that

i) when $\alpha\to 0$, (\ref{t}) is the partition
function per unit volume of a charged particle in the average magnetic field,
with a Landau spectrum shited by
$<\omega_c>$
\be \label{alpha}{Z}={1\over 4\pi}{<B>\over \sinh
t<B>/2}e^{-t<B>/2}={1\over 2\pi t}(1-\sum_{n=0}^{\infty}{\zeta(-n)\over
n!}(-<B>t)^{n+1})\ee
(again this follows directly from general properties of Brownian
curves -note also that the attractive-core case would be obtained simply
by $<B>\to-<B>$ in (\ref{alpha})).

ii)  from the $t$
expansion of the specific heat $c=c_ot^2{d^2\ln Z\over dt^2}$ that
the $\alpha_o$ value is attained when
\be\label{attain} <(S+iA)^2>=<S>^2\ee

It follows that by a diagrammatic expansion in $\alpha$ of quantities
such as
${Z_1\over Z_0}-1$, $ {Z_2\over Z_0}-1-2({Z_1\over Z_0}-1)$, $\cdots$,
one should recover

i) the leading $\alpha$ behavior (\ref{alpha})

ii) the critical $\alpha_o$ value by finding the zeroes of
$({Z_1\over Z_0}-1)^2={Z_2\over Z_0}-1-2({Z_1\over Z_0}-1)$

The diagrammatic expansion of $Z_N$ is performed by usual perturbative
methods for the Hamiltonian (\ref{101}).
One uses a non perturbed basis of free thermal propagators
$G_{t}(\vec r,\vec r')={1\over
2\pi t}\exp(-(\vec r-\vec r')^2/2t)$. In the computation of the
average $N$ impurity
 partition function, one encounters, at a given order in $\alpha$,
volume divergences   from the $dz_id\bar z_idz_jd\bar z_jd... $
space integrals over the impurity locations $\vec r_i, \vec r_j, \cdots$, if
computed directly in the thermodynamic limit. Thus, in
principle, the need
of a certain regularization prescription, as for example
a harmonic regularization.
However, in our case, one can circumvent this difficulty by
computing, rather than average partition functions $Tr \exp(- t H_N)$,
the
average of the
 thermal propagators
 $G_N(\vec r_o,\vec r_o)\equiv<\vec r_o|\exp(-t {\tilde{H}}_N)|\vec r_o>$ from
and to a given point $\vec r_o$.
The actual average partition functions are by definition
space integral over $\vec r_o$ of these propagators. Yet, averaging
over disorder has to be made. If it is done before the final $\vec r_o$ space
integration, it follows that
$<G_N(\vec r_o,\vec r_o)>$ does not depend on $\vec r_o$ anymore.
Therefore, the final space integration becomes trivial, since it amounts
to multiply by $V$ the coinciding point propagator. In this
computational scheme, which relies on the crucial fact that averaging on
the disorder has to be made, there is no need for a particular
regularization procedure, since the infinite volume factors out
trivially
in the last space integral.

At a given order $\alpha^n$, a diagram has $n$ vertex, and $m\le n$
impurities: it contributes at order $\alpha^n \rho^m$ in the
concentration expansion. The diagrams which contribute to the $<B>$ average
partition function are necessarily such that $m=n$, and corrections to
the average field limit are built by diagrams with $m<n$.
We have analytically computed the leading $<B>$ diagrams up to order
$\alpha^4\rho^4$, and at order $\rho^2$ -the two impurity case-, the
corrections to the leading $\alpha^2\rho^2$ digram up to order
$\alpha^4$. The diagrammatic expansion results are:

\be\label{diagr} {Z_1\over Z_0}-1={1\over VZ_o}{\alpha(\alpha-1)\over 2}\ee

\be  \label{diag}{Z_2\over Z_0}-1-2({Z_1\over Z_0}-1)={1\over
(VZ_o)^2}\big({1\over 6}
\alpha^2 + 0\alpha^3+
({1\over 24}-{7\over 16}\zeta(3))\alpha^4 +\cdots\big)\ee
where ${1\over 24}-{7\over 16}\zeta(3)=-0.48423323$,
\be  \label{diagrr}{Z_3\over Z_0}-3{Z_2\over Z_0}+3{Z_1\over Z_0}-1={1\over
(VZ_o)^3}
\big(0\alpha^3 + \cdots\big)\ee

\be  \label{diagrrr}{Z_4\over Z_0}-4{Z_3\over Z_0}
+6{Z_2\over Z_0}-4{Z_1\over Z_0}+1={1\over (VZ_o)^4}\big(-{1\over
30}\alpha^4+\cdots  \big)\ee
We see that at leading order $\alpha^n\rho^n$,
(\ref{diagr}, \ref{diag}, \ref{diagrr}, \ref{diagrrr}) indeed reproduce
the average magnetic field expansion (\ref{alpha}).

We are now in position to determine the value $\alpha_o$ at which the
specific heat transition occurs, just by considering
(\ref{diagr},\ref{diag}) on the one hand, (\ref{relat},\ref{attain}) on the
other hand. Before doing so, the
diagrammatic expansion (\ref{diag}) should be first completed to obtain a
$\alpha(1-\alpha)$
polynomial,
since this should be so. One gets
\be  \label{diagra}{Z_2\over Z_0}-1-2({Z_1\over Z_0}-1)={1\over
(VZ_o)^2}\bigg(
{1\over 6}[\alpha(1-\alpha)]^2 + {1\over
3}[\alpha(1-\alpha)]^3+{7\over 8}(1-{1\over
2}\zeta(3))[\alpha(1-\alpha)]^4 + O\big([\alpha(1-\alpha)]^5\big)\bigg)\ee
The resulting polynomial equation in $\alpha$ writes
\be {\big({\alpha(1-\alpha)\over 2}\big)}^2=
{1\over 6}[\alpha(1-\alpha)]^2 + {1\over
3}[\alpha(1-\alpha)]^3+{7\over 8}(1-{1\over
2}\zeta(3))[\alpha(1-\alpha)]^4 \ee
In the
interval $\alpha\in [0,1/2]$, this equation has the desired root
$\alpha_o=0.29$.

One can go a bit further, and compare
the diagrammatic expansion with the numerical estimations for
$(<S^2>_{\{C\}}-
<S>^2_{\{C\}})-<A^2>_{\{C\}})$. In Fig. 4, not only
the three curves - a) and b) are numerical simulations, c) is the
diagrammatic expansion- have
the same intercept with the horizontal axis (thus the same
$\alpha_o\simeq\alpha_o^{num}$
value), but also they  exhibit the same qualitative behavior.
Differences
are due to the fact that actual
winding properties for true Brownian walks are difficult to reach from
finite length random walks, and to the fact that the perturbative
expansion in $\alpha(1-\alpha)$ is by definition incomplete. Still, it is
remarkable that both approaches yield the same value for $\alpha_o$,
altogether with the same $\alpha$ qualitative behavior.
A convergence between both approaches is expected.

An interesting consequence of the discussion above concerns the average
density of states for the one impurity problem, two impurities problem,
etc.
In the simple one impurity case, averaging over disorder is trivial,
since the Aharonov-Bohm low energy depletion of states does not
depend on the position of the vortex. One simply recovers the standard
result  \cite{deplet}
\be <\rho_1(E)-\rho_o(E)>=\rho_1(E)-\rho_o(E)=
{\alpha(\alpha-1)\over 2}\delta(E)\ee
In the two-impurities case, things become highly non trivial, but averaging
has allowed for the diagrammatic expansion result (\ref{diag}). One deduces
that
\begin{eqnarray}
<\rho_2(E)-\rho_o(E)>-&2&<\rho_1(E)-\rho_o(E)>=
{2\pi\over V}
\bigg(
{1\over 6}[\alpha(1-\alpha)]^2 + {1\over
3}[\alpha(1-\alpha)]^3\nonumber\\&+&{7\over 8}(1-{1\over
2}\zeta(3))[\alpha(1-\alpha)]^4 + \dots\bigg)
 \delta'(E)\end{eqnarray}
\be
<\rho_3(E)-\rho_o(E)>-3<\rho_2(E)-\rho_o(E)>+3<\rho_1(E)-\rho_o(E)>
=  \big({2\pi\over V}\big)^2
(0\alpha^3+\cdots)\delta^{''}(E)\ee
\begin{eqnarray}
<\rho_4(E)-\rho_o(E)>-4<\rho_3(E)-\rho_o(E)>&+&6<\rho_2(E)-\rho_o(E)>
-4<\rho_1(E)-\rho_o(E)>=  \nonumber\\\big({2\pi\over V}\big)^3
(-{1\over 30}\alpha^4&+&\cdots)\delta^{'''}(E)\end{eqnarray}
etc...
At leading order $\rho^n\alpha^n$, summing up all leading
contributions from the one, two, $\cdots$ impurity cases, leads to
\begin{eqnarray}<\rho(E)-\rho_o(E)>&=&{V<B>\over 2\pi}\bigg(-{1\over
2}\delta(E)+{<B>\over 2!}{1\over 6}\delta'(E)\nonumber\\
& & \quad+{<B>^2\over
3!}0\delta^{''}(E)+{<B>^3\over 4!}(-{1\over
30})\delta^{'''}(E)+\cdots\bigg)\nonumber\\
&=&{V<B>\over
2\pi}\sum_{n=1}^{\infty}\delta(E-n<B>)-\rho_o(E)=
\rho_L(E)-\rho_o(E)\end{eqnarray}
which reproduces, as expected,  the
shifted Landau density of
states $\rho_L(E)-\rho_o(E)$.

{\bf IV.c Starting from the Landau basis:}

Perturbative computations can also be done by starting directly from the
Hamiltonian (\ref{101}). The expansion of the average
partition function will be modified
in the following way:

-  propagators to be used are Landau propagators in the average
magnetic field
$$G_{t}(r,r')={<\omega_c>\over 2\pi\sinh t<\omega_c>}
e^{\displaystyle -{<\omega_c>\over 2\sinh t<\omega_c>}[\vert z-z'\vert^2\cosh t
<\omega_c>
+\sinh t <\omega_c>(z\bar z'-\bar z z')]}$$

instead of free propagators.

- the vertex to be used is
$$-2\alpha{1\over\bar z-\bar z_i'}(\partial_z-{1\over 2}<\omega_c>\bar z)
+2\alpha{\pi\rho\over N}(z\partial_z-{1\over 2}<\omega_c>z\bar z)$$
instead of
$$-2{\alpha}{\partial_z\over \bar z-\bar z_i}$$

One has already noted that any diagram with isolated impurities only, i.e. of
the
$\alpha^n\rho^n$ type,
contributes to the average magnetic field partition function.
It follows that, by definition, it vanishes in the present formulation
since the average magnetic field contributions are  ab initio incorporated
in the Landau propagator.
It is possible to show that for any subleading
diagrams $\alpha^n\rho^m, m<n$, the vertex reduces to
$$-2\alpha{1\over\bar z-\bar z_i}(\partial_z-{1\over 2}<\omega_c>\bar z)$$

Any  such subleading diagram automatically incorporates the corresponding
diagram in the free
propagator approach, plus all diagrams
deduced from it by adding an arbitrary
number of isolated impurity lines, i.e. diagrams of the type $\alpha^n
\rho^m(1+\alpha\rho+\alpha^2\rho^2+\cdots), m<n$.

{\bf V. Conclusion}

A more precise value of $\alpha_c$ and a better
understanding of what is exactly
happening at the transition is still missing. How Landau levels
actually appear when $\alpha$ continuously decreases
from $\alpha_c$ to 0? What is the exact nature of the transition?

Another important issue concerns the conductivity and localisation
properties of the test particle in the random system of magnetic
impurities. In particular, one would like to have information on
the  Hall conductance in the impurity system.
This question could be partially adressed by considering the
average
persistent currents
due to the vortex distribution\cite{persis}. One has been able to get
perturbative expansion in $\alpha$ for the one, two, $\cdots$ impurity average
densities of states. In principle, one can deduce \cite{persis} from these
density of states the
average persistent currents for one
vortex, two vortices, $\cdots$. In the limit $\alpha\to 0$, at leading
order $\alpha^n\rho^n$,
diamagnetic persistent currents for a constant magnetic field should be
recovered
by
summing up all contributions coming from the $N$ impurities persistent
currents.

Figure captions :

Figure 1: The average level density of states $<\rho(E)>$
as a function of the
variable ${E/\rho}$ for $\alpha=1/2$
exhibits a Lifshitz tail at the bottom of the spectrum.

Figure 2: The variances of the random variables $S$ and $A$ obtained by
simulations (2000 closed random walks of 100000 steps) are plotted as a
function of $\alpha$. The intersection of the two curves  at
$\alpha_o^{num}=0.28$ determines a change of the specific heat behavior at high
temperature.

Figure 3: The average level density of states, determined by numerical
simulations,  as a function of the
variable $E/\rho$ for different $\alpha$ values. When
$\alpha<\alpha_c\simeq 0.35$, the
density of states is no more  monotonic and oscillations appear.

Figure 4: The difference between the variances of $S$ and $A$ as a
function of $\alpha$, determined by numerical simulations for a) 3000
closed random walks of 400 steps, b) 2000 closed random walks pf 100000
steps, and also by the diagrammatic expansion to fourth order in
$\alpha$. The three curves intercept the horizontal axis at
$\alpha\simeq 0.28-0.30$. The origin of the quantitative differences are
explained in the text.

\end{document}